\documentclass[a4paper,11pt]{article}
\pdfoutput=1 

\usepackage{jcappub} 

\usepackage[T1]{fontenc} 

\usepackage{latexsym}
\usepackage{amsfonts}
\usepackage{amsmath}
\usepackage{amssymb}
\usepackage{graphicx,color}
\usepackage{float}
\usepackage{hyperref}
\usepackage{subfigure}
\usepackage{dcolumn}
\usepackage{soul}
\usepackage{ulem}
\usepackage{verbatim}
\usepackage{xcolor}

\arxivnumber{2504.15606}

\title{Warm multi-natural inflation}

\author[a,1]{Asuka Ito \note{Corresponding author.}}
\author[b]{and Rudnei O. Ramos}

\affiliation[a]{Department of Physics, Kobe University, Kobe 657-8501, Japan}
\affiliation[b]{Departamento de F\'{\i}sica Te\'orica, Universidade do Estado do Rio de Janeiro, 20550-013 Rio
de Janeiro, RJ, Brazil}

\emailAdd{asuka@phys.sci.kobe-u.ac.jp}
\emailAdd{rudnei@uerj.br}

\abstract{

Multi-natural inflation is studied in the context of warm inflation.
We study the warm multi-natural inflation scenario with both linear and cubic temperature-dependent
dissipation coefficients.
The model is motivated by axion-like inflation models with coupling to non-Abelian gauge fields through a
dimension-five coupling and dissipation originating from sphaleron decay in a thermal bath.
Both cases of dissipation coefficients can be compatible with current observations.
In the case of the cubic dissipation coefficient, we find that 
the curvature perturbation starts to grow suddenly when a transition from a weak dissipation
to a strong dissipation regime occurs at the later stage of the inflation.
We also show that such rapid growth of the curvature perturbation on small scales 
gives rise to abundant scalar induced gravitational waves, which may be detectable with future
gravitational wave detectors such as DECIGO and ET. On the other hand, there are also other parameter
regions of the model, in the warm inflation regime of weak to strong dissipation and with
sub-Planckian axion decay constant, that can lead to overproduction of
primordial black holes on small scales, which are constrained by nucleosynthesis  bounds,
thus ruling out the model in this region of parameters.

}

\begin{document}

\maketitle

\tableofcontents
\section{Introduction}

Inflationary cosmology has profoundly shaped our understanding of the early universe, providing robust solutions to the horizon and flatness problems while seeding primordial density fluctuations. Inflation models can be classified into two main categories: {\it cold inflation} (CI) and {\it warm inflation} (WI) models.
In the traditional CI model, a supercooled universe is followed by a post-inflationary reheating phase to generate the hot Big Bang. On the other hand, WI~\cite{Berera:1995ie} proposes a scenario in which the inflaton field continuously dissipates energy into a thermal radiation bath during inflation, rather than after. This paradigm avoids supercooling, naturally sustains slow-roll through dissipative dynamics, and seamlessly transitions to a radiation-dominated era without a distinct reheating phase~\cite{Berera:1996nv,Berera:1998gx,Berera:1996fm}. Moreover, warm inflation's thermal fluctuations offer distinct observational signatures, such as modified primordial power spectra and non-Gaussianities, testable against cosmic microwave background (CMB) data~\cite{Bastero-Gil:2014jsa,Bastero-Gil:2014raa,Benetti:2016jhf,Bastero-Gil:2017wwl,Arya:2017zlb} (for recent work confronting different WI models
with data, see, e.g., refs.~\cite{Kumar:2024hju,Santos:2024pix,Santos:2024plu,DAgostino:2024nni,Berera:2025vsu}). Its compatibility with particle physics models, where dissipation arises naturally from interactions in well-motivated quantum field theory models~\cite{Berera:2008ar,BasteroGil:2010pb,BasteroGil:2012cm},
further motivates its study. The recent advances in WI (see, for example~\cite{Kamali:2023lzq} and references therein), underscore WI as a compelling bridge between early-universe cosmology and high-energy theory.

The model we study in this paper is an inflationary scenario involving an axion-like field with two cosine potentials, known as multi-natural inflation~\cite{Czerny:2014wza}. 
This setup is motivated by a broken global U(1) symmetry, in which a Peccei–Quinn field couples to two distinct sets of quarks, each charged under different non-Abelian gauge groups.
We implement the multi-natural inflation model in the context of WI.
In WI, the dissipation coefficient, denoted as $\Upsilon$, quantifies the rate at which the inflaton field transfers energy to a thermal bath, sustaining radiation production during inflation. Its functional form, which usually can be a function of the temperature $T$ of the thermal radiation bath and the background
inflaton field $\phi$, depends on the underlying particle physics interactions and the regime of the inflaton's coupling to other fields (for a recent review, see also ref.~\cite{Kamali:2023lzq}). 
Well-studied cases of the dissipation coefficient in WI
include functional dependencies that can be linear or cubic in the temperature.  
These cases emerge for instance in cases where the inflaton
is a pseudo-Goldstone boson~\cite{Bastero-Gil:2016qru,Bastero-Gil:2019gao,Berghaus:2019whh,Laine:2021ego}. 
Hence, these forms of dissipation
coefficients studied here suit to the type of model we consider, since the multi-natural inflation model is an axion-like type of model.

By implementing the multi-natural inflation model in the context of WI, we show that
in the case of the cubic dissipation coefficient, the scalar perturbations
start to increase suddenly when a transition from a weak dissipation regime to a strong dissipation regime occurs at the later stage of the inflation.
The tensor parts of perturbations cannot interact with scalar perturbations at linear order if the background spacetime is isotropic, though they can interact with 
each other nonlinearly~\cite{Baumann:2007zm,Ananda:2006af}.
This indicates that scalar-induced tensor perturbations can be important only when
the scalar perturbations take large values close to unity.
We show that large scalar perturbations on small scales created during 
the warm multi-natural inflation with the cubic dissipation coefficient can 
give rise to abundant gravitational waves.
Depending on the inflation energy scale, this gravitational wave spectrum can have a peak in the high-frequency regime around MHz, whose detection methods have been explored  
intensively~\cite{Aggarwal:2025noe,Ito:2023nkq,Ito:2023bnu,Ito:2023fcr,Ito:2022rxn,Ito:2020wxi,Ito:2019wcb}.
Moreover, its low-frequency tail may be detectable with future gravitational wave 
detectors such as DECIGO and ET.
Therefore, the warm multi-natural inflation is testable with gravitational wave observations
in principle.

The organization of the paper is as follows.
In sec.~\ref{sec2}, the WI scenario with a multi-natural axion potential 
is explained.
In sec.~\ref{sec3}, background equations of motion and perturbative equations 
are investigated in the warm multi-natural inflation with the two forms of dissipation coefficient
considered in this paper. We derive approximate analytical expressions which helps to understand
the behavior seen in the background dynamics and also the perturbations get affected in the model
under study.
In sec.~\ref{sec4}, we numerically solve the equations of motion in the cases of 
the linear and cubic dissipation coefficients.
The behavior of the dynamics that emerges in both cases are discussed. We also analyze the scalar 
power spectrum obtained in both cases. Here, we also investigate the dependence of the relevant background and perturbation
quantities on the axion decay constant in the multi-natural inflaton
potential.
In sec.~\ref{GWsection}, we evaluate the scalar-induced gravitational waves on small scales.
The focus is given in the case of the cubic dissipation coefficient, where the most notable results appear.
Compatibility of our results with the bounds on the scalar power spectrum on small scales 
from light primordial black hole evaporation is also discussed.
In sec.~\ref{conclusions}, we give our conclusions and final remarks.
An appendix is also included to show and explain some key results that we have found.

\section{warm multi-natural inflation}
\label{sec2}

In the present implementation  of WI, we work with an
axion-like  field $\phi$ that plays
the role of the inflaton and which is coupled to a non-Abelian $SU(N_c)$ gauge
field $A_\mu$ with the standard dimension five interaction, 
\begin{eqnarray}
{\cal L}_{\rm int} &=& - \frac{g^2}{32\pi^2 f}\phi  F_{\mu \nu}^c \tilde{F}^{c\,
    \mu \nu},
\label{Lint}
\end{eqnarray}
where $c=1,\ldots, N_c^2-1$ is the group index, $F_{\mu \nu}^a=
\partial_\mu A_\nu^a - \partial_\nu A_\mu^a +g \epsilon^{abc} A_\mu^b
A_\nu^c$ is the  gauge field tensor, $\tilde{F}^{c\,\mu \nu}$ is its
dual,
\begin{equation}
\tilde{F}^{c\,\mu \nu}=\frac{\epsilon^{\mu \nu \rho \sigma}}{2 a^3(t)}
F_{\rho \sigma}^c,
\end{equation}
$a(t)$ is the scale factor and $\epsilon^{\mu \nu \rho \sigma}$ is the
usual Minkowski-like completely antisymmetrical tensor. 

The interaction (\ref{Lint}) is known to lead to a dissipation coefficient,
which comes from nonperturbative sphaleron rate transitions at finite
temperature in the gauge vacua,
which is given
by~\cite{Berghaus:2019whh,Laine:2021ego} 
\begin{equation}
\Upsilon_{\rm sph}(T)= \kappa \frac{T^3}{f^2},
\label{UpscubicT}
\end{equation}
where $\kappa$ is given by
\begin{equation}
\kappa \simeq 1.2 \pi\frac{g^4 (g^2 N_c)^3
  (N_c^2-1)}{(64\pi^3)^2}\left[\ln\left(\frac{m_D}{\gamma}\right)+3.041\right],
\label{kappa}
\end{equation}
where $m_D^2=g^2 N_c T^2/3$ is the Debye mass squared of the
Yang-Mills plasma and $\gamma$ is given by the solution
of~\cite{Moore:2010jd}
\begin{equation}
\gamma=\frac{g^2 N_c T}{4 \pi}
\left[\ln\left(\frac{m_D}{\gamma}\right)+3.041\right].
\end{equation}
This gives for $\kappa$ the result
\begin{eqnarray}
\kappa &= & 0.3 \alpha_{g}^5 N_c^3 (N_c^2-1) W\left(
e^{3.041}\sqrt{\frac{4\pi}{3\alpha_{g} N_c} }\right),
\label{kappa2}
\end{eqnarray}
where $\alpha_{g}=g^2/(4\pi)$ is the fine structure Yang-Mills coupling and $W(x)$ is the
Lambert function, given by the principal solution of $x=w e^w$. If we assume, for example,
$N_c=3$ and  $\alpha_{g}=0.1$, this gives $\kappa \simeq
2.07 \times 10^{-3}$.

The result eq.~(\ref{UpscubicT}) applies in the absence of fermions. By having fermions
coupled to the gauge field, the Lagrangian density for the bath fields ($A_\mu^c,\; \psi,; \bar \psi$)
becomes 
\begin{equation}
{\cal L}_{\rm bath}=-\frac{1}{4} F_{\mu \nu}^c F^{c\,\mu \nu}
+\bar{\psi} (i \slash \!\!\!\! D-m)\psi,
\label{Lbath}
\end{equation}
where $\slash \!\!\!\! D = \gamma^\mu D_\mu=\gamma^\mu (\partial_\mu + i g A_\mu^c {\cal T}^c)$,
with ${\cal T}^c$ the generators of the non-Abelian gauge group.
The presence of fermions tends to suppress the sphaleron dissipation in the case of 
chiral symmetry breaking ($m\neq 0$) and can even make it completely vanish, for
massless fermions ($m=0$). The effective dissipation in the presence of fermions
become~\cite{Berghaus:2020ekh,Drewes:2023khq,Berghaus:2024zfg}
\begin{equation}
\Upsilon_{\rm eff} = \Upsilon_{\rm sph} \left( 
\frac{\Gamma_{\rm ch}}{\Gamma_{\rm ch}+ 2 \frac{f^2}{T^2} \Upsilon_{\rm sph} } \right).
\label{Upseff}
\end{equation}
where $\Upsilon_{\rm sph}$ is the pure gauge result eq.~(\ref{UpscubicT}) and
$\Gamma_{\rm ch}$ is given by
\begin{equation}
\Gamma_{\rm ch} \sim  N_c \alpha_{g} \frac{m^2}{T},
\label{Gammach}
\end{equation}
where we are assuming that the chiral decay processes
happen at a sufficiently large rate\footnote{When $\Gamma_{\rm ch} \ll H$, the effective dissipation
eq.~(\ref{Upseff}) changes~\cite{Berghaus:2025dqi,ORamos:2025uqs}, but it still falls in an intermediate class
of dissipation in between the two cases we consider here.}, $\Gamma_{\rm ch}  > H$, which can always be arranged
by suitable choices of parameter, e.g. $\alpha_g$ and $m$.
{}From eq.~(\ref{Upseff}) we see that in the chiral limit $m\to 0$ the dissipation
vanishes, for heavy fermions, $m \gg T,\,f$, we recover the result given by
eq.~(\ref{UpscubicT}), while for light fermions the dissipation becomes linearly
dependent on the temperature,
\begin{equation}
\Upsilon_{\rm eff} \simeq  r N_c \alpha_{g} \frac{m^2}{2f^2}T.
\label{UpslinearT}
\end{equation}
The compatibility of WI in the context of QCD like values for the fermions (quarks) have been 
explicitly verified recently by the authors of ref.~\cite{Berghaus:2024zfg}.

As explicitly verified in many recent works
(see, for example refs.~\cite{Laine:2021ego,DeRocco:2021rzv,Das:2022ubr,Montefalcone:2022jfw,Berghaus:2024zfg,ORamos:2025uqs}),
dissipation of the form of eq.~(\ref{UpscubicT}) is able to lead to a consistent WI regime
for a variety of inflaton potentials. The same is true for $\Upsilon \propto T$, as explicitly verified
in many previous references~\cite{Bastero-Gil:2016qru,Benetti:2016jhf,Ballesteros:2023dno,Santos:2024pix}.
Yet, as far as standard axion-like cosine potentials are
concerned, it is still a challenge to produce regimes of either cold or warm inflation for which
the axion-decay constant is sub-Planckian, $f_a < M_{\rm Pl}$ (see, e.g., refs.~\cite{Montefalcone:2022jfw,Zell:2024vfn}
for some recent work in the context of axion WI). 
To overcome these issues with the standard axion-like cosine potential, in this paper we will 
work with a multi-natural axion like potential given by~\cite{Czerny:2014wza}
\begin{equation}
V(\phi) = V_0\left[ \cos\left( \frac{\phi}{f_a} + \theta\right) - \frac{\kappa}{n^2}
\cos\left( n\frac{\phi}{f_a}\right) \right] +C,
\label{pot}
\end{equation}
where $C$ is a constant, which is adjusted such that $V(\phi_{\rm min})=0$
at its minimum. In the above potential $f_a$ is the axion decay constant, 
while $\kappa$, $\theta$ and $n$ are dimensionless constants.
Here, we study this model in the context of WI with both cases of dissipation coefficients, given by
eqs.~(\ref{UpscubicT}) and (\ref{UpslinearT}).
{}Following~\cite{Daido:2017wwb}, we assume that $\kappa =1$, $n$ is an odd integer and $\theta \ll 1$.
The motivation for using odd values stems from the fact that, in this case, the second derivatives of the potential at both its maximum and minimum are equal in magnitude but opposite in sign. When $n$ is odd and increases, both the top and bottom of the potential become flatter at an equal rate.
In multi-natural inflation, this offers several advantages. Firstly, it facilitates achieving inflation along the flatter top. Secondly, by appropriately setting the phase $\theta$ to be very small, the inflaton can be arranged to have a very small mass when it oscillates around the minimum. This allows the inflaton to be long-lived and potentially contribute to dark matter. Thus, by carefully tuning the potential's parameters, a sub-Planckian axion decay constant can be achieved while satisfying observational constraints for inflation (e.g. from Planck~\cite{Planck:2018jri}), and an inflaton remnant could also serve as dark matter. While the first advantage (achieving inflation) could also be arranged for an even value of $n$ \footnote{Note that even values for $n$, e.g. $n=2$, is also motivated from
QCD axions~\cite{GrillidiCortona:2015jxo}.}, the second (small inflaton mass and dark matter contribution) would not be possible.

\section{Background dynamics and perturbation quantities in WI}
\label{sec3}

The relevant background equations in WI are the evolution equations for the inflaton field
and radiation energy density,
\begin{eqnarray}\label{eqphi}
&&\!\!\!\!\!\!\!\!\!\!\! \!\!\!\!  \ddot{\phi}+3H\dot{\phi}+V_{,\phi}
    +\Upsilon\dot{\phi}=0,
\\ && \!\!\!\!\!\!\!\!\!\!\!
\!\!\!\!
\dot{\rho}_{r}+4H\rho_{r}=\Upsilon\dot{\phi}^2.
\label{eqrhor}
  \end{eqnarray}

{}For a thermalized radiation bath, we have that $\rho_r$ is
related to the temperature of the thermal bath as
 \begin{eqnarray}
\rho_{r}=C_r T^4,
\label{rhor}
 \end{eqnarray}
where $C_r = g_* \pi^2/30$ and $g_*$ denotes the radiation bath
degrees of freedom.  In the present work we assume that the radiation
bath is constituted primarily by the gauge field fluctuations, which
then gives $g_*=2(N_c^2-1)$ for the case of the  (massless) $SU(N_c)$
Yang-Mills fields\footnote{When including also the inflaton
fluctuations as thermalized, then $g_*=2N_c^2-1$.}.

In WI, the slow-roll variables become~\cite{Kamali:2023lzq}
 \begin{eqnarray}\label{Slow-roll}
&& \epsilon_H=-\frac{\dot{H}}{H^2}\simeq \frac{\epsilon_V}{1+Q},
\\
&& \eta=\frac{\eta_V}{1+Q},
 \end{eqnarray}
where $Q= \Upsilon/(3H)$ denotes the dissipation ratio in WI,
while $\epsilon_V$ and $\eta_V$ are the standard cold inflation slow-roll coefficients,
defined, respectively, as
\begin{eqnarray}
&&\epsilon_V= \frac{M_{\rm Pl}^2}{2} \left(\frac{V_{,\phi}}{V}\right)^2,
\label{epsV}
\\
&&\eta_V= M_{\rm Pl}^2 \frac{V_{,\phi\phi}}{V}.
\label{etaV}
\end{eqnarray}

The general expression for the scalar of curvature
perturbation spectrum in WI is~\cite{Kamali:2023lzq}
\begin{align}
P_{\mathcal{R}}\simeq \left(\frac{H_{*}^2}{2 \pi \dot\phi_{*}}\right)^2
 \left(1+2n_{*} + \frac{2\sqrt{3}\pi
  Q_{*}}{\sqrt{3+4\pi Q_*}}{T_{*}\over H_{*}}\right) G(Q_*),
\label{powers}
\end{align}
where $n_*$ in the above expression denotes the possible statistical distribution for the inflaton
due to the presence of the radiation bath, while the multiplicative
function of the dissipation coefficient, $G(Q)$, in
eq.~(\ref{powers}) depends on the explicit form of the dissipation coefficient
$\Upsilon$, e.g. on how $\Upsilon$ depends on the temperature and inflaton
amplitude\footnote{Note that while $G(Q)$ depends mostly on the form of the dissipation
coefficient, it is weakly dependent on the form of the inflaton potential~\cite{Montefalcone:2023pvh,Rodrigues:2025neh}.}.
The subindex $*$ in the quantities in eq.~(\ref{powers}) means that
they are evaluated at the Hubble crossing time, i.e.,
when $k=a H$.

The behavior of the power spectrum in WI is essentially dependent on the evolution of $Q$ and $T/H$, 
which in turn depends on the form of the dissipation coefficient $\Upsilon$ and
the inflaton potential. Parameterizing $\Upsilon$ for example like
\begin{equation}
\Upsilon(\phi, T) = C_{\Upsilon} T^{c} \phi^p M_{\rm Pl}^{1-p-c},
\label{UpsTphi}
\end{equation}
we have~\cite{Das:2020lut,Das:2022ubr}
\begin{eqnarray}
\frac{d \ln Q}{d N} & =& C_{Q}^{-1}
 \left[(2c+ 4) \epsilon_{V} - 2c\eta_{V} - 4 p
   \kappa_{V}\right], \nonumber \\
\label{b3}
\\ 
\frac{d\ln (T/H)}{d N} &=& C_{Q}^{-1}
 \left[\frac{7-c + (5+c)Q}{1+Q} \epsilon _{V} - 2\eta_{V} -
   \frac{1-Q}{1+Q} p \kappa_{V}\right], \label{b4}
\end{eqnarray}
where $\kappa_{V} = M^2_{Pl} {V_{\phi}}/{(\phi V)}$ and $C_{Q} = 4-c + (4+c) Q $  
is a positive quantity, since Q is always positive and $-4<c<4$ is due to stability
conditions in WI~\cite{Moss:2008yb,Bastero-Gil:2012vuu,delCampo:2010by}.  
The eqs.~(\ref{b3}) and (\ref{b4}) show that $Q$ and $T/H$ will grow during
WI for $c>0$ and for sufficiently concave potentials (where $\eta_V <0$ during
slow roll). Under these circumstances, the power spectrum $P_{\mathcal{R}}$
can increase during the WI dynamics. This effect has been explored recently
in connection with the production of primordial black holes in WI~\cite{Arya:2019wck,Bastero-Gil:2021fac,Correa:2022ngq}
(for a slight different context, see also ref.~\cite{Ferraz:2024bvd} which also employs WI
in the context of primordial black hole and gravitational wave generation).
We will explore this ability of the power spectrum to grow during WI in connection to the
model studied here.

In WI we need to enforce several constraining conditions, such as to avoid a strong backreaction from the gauge field~\cite{Kamali:2024qme}, enforce
perturbativity ($\alpha_g <1$) such that
the calculation leading to eq.~(\ref{UpscubicT}) remains valid, etc. These conditions
typically lead to a hierarchy of energy scales~\cite{Berghaus:2019whh,Laine:2021ego,DeRocco:2021rzv,Kamali:2024qme},
$H<T<f<f_a$. In all of our examples, we observe that all these conditions remain satisfied.

\begin{center}
\begin{figure*}[!bth]
\subfigure[]{\includegraphics[width=7.cm]{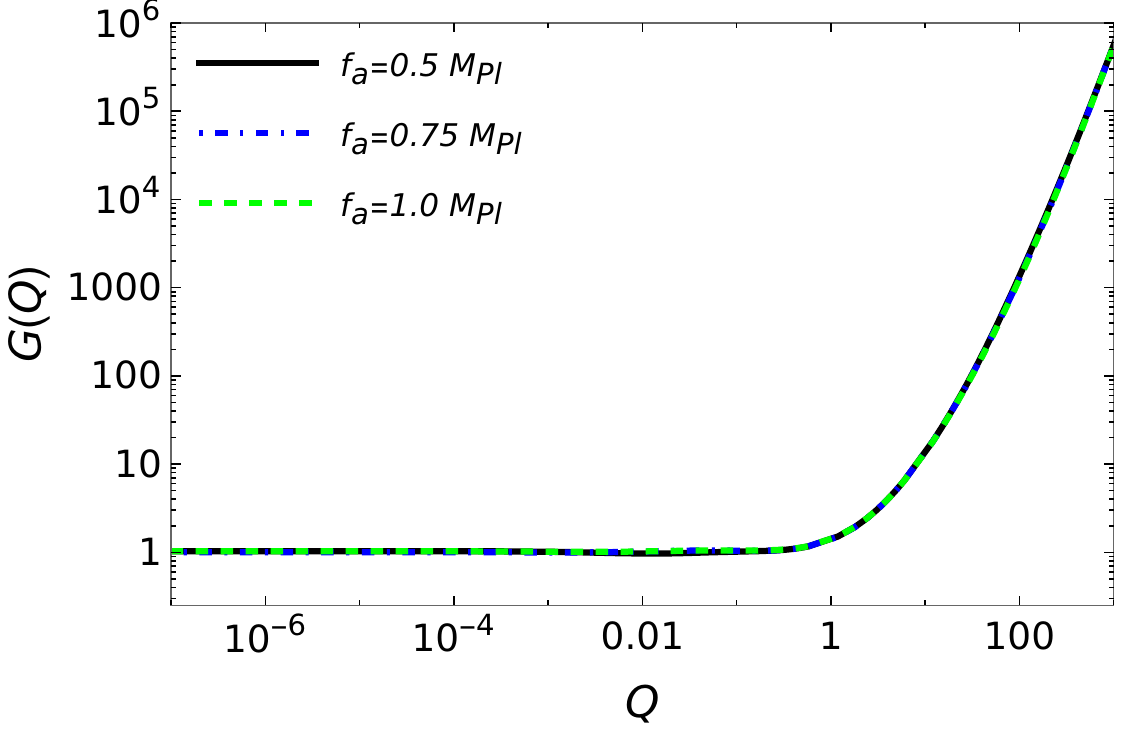}}
\subfigure[]{\includegraphics[width=7.cm]{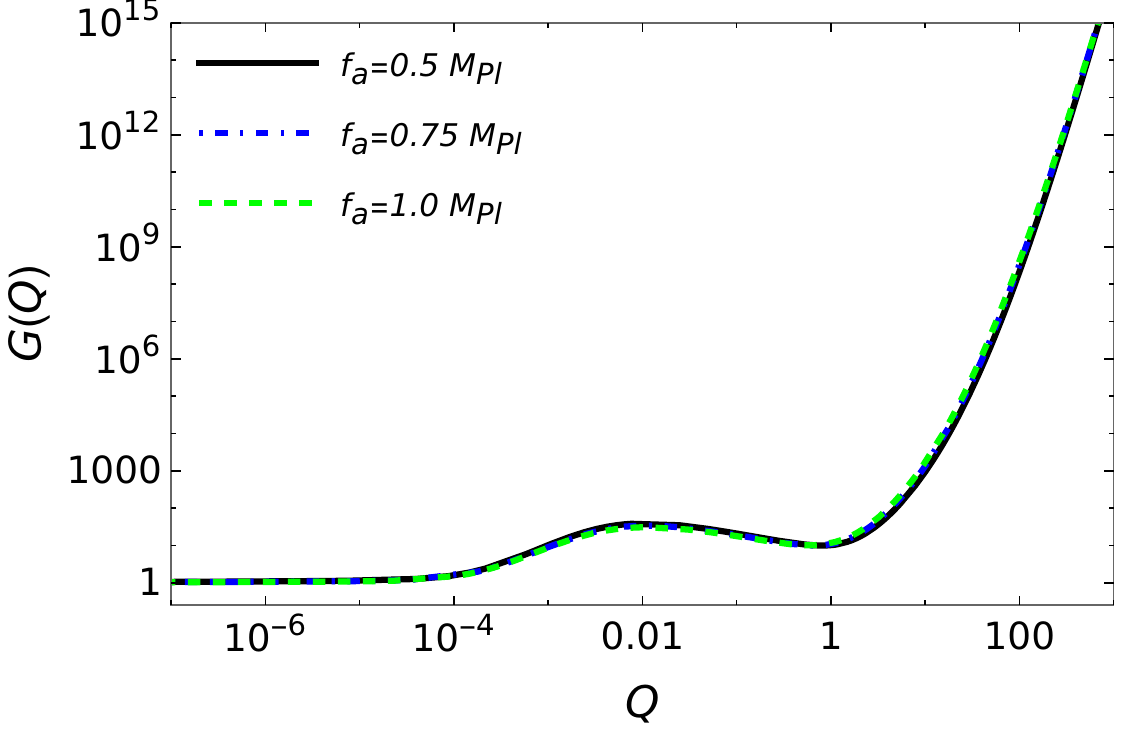}}
\caption{The $G(Q)$ function generated through  
\texttt{WI2easy}~\cite{Rodrigues:2025neh} for a dissipation coefficient eq.~(\ref{UpsTphi}) for $c=1,p=0$ (panel a) and for $c=3,p=0$ (panel b) and
obtained for three different values for the axion decay for the potential (\ref{pot})
(considering $n=3$, $\kappa=1$ and $\theta$ as given in tables~\ref{tab1} and \ref{tab2}
for each dissipation case, respectively). The nonthermal case $n_*=0$ is considered.}
\label{fig1}
\end{figure*}
\end{center}

In using eq.~(\ref{powers}) for the scalar power spectrum in WI, we also need
to check whether the inflaton will be thermalized with the thermal bath or not~\cite{Bastero-Gil:2017yzb}.
This determines, in particular, the presence or absence of the term $n_*$ in eq.~(\ref{powers}).
{}For an interaction of the form of eq.~(\ref{Lint}) the dominant contribution for the thermalization
of the inflaton (axion) with the gauge fields (thermal bath) comes from the scattering processes
such as $\phi+A_\mu \leftrightarrows A_\mu + A_\mu$, which occurs at a rate~\cite{Masso:2002np}
\footnote{The contributions from processes involving fermions of the thermal bath
can also be included, but they contribute  subdominantly  to the rate when compared to the one due 
to the pure gauge field~\cite{Masso:2002np}.}
\begin{equation}
\Gamma_{\phi+A_\mu \leftrightarrows A_\mu + A_\mu} \sim \frac{\alpha_g^3 T^3}{32 \pi f^2}.
\label{rate}
\end{equation}
Throughout our analysis we assume $n_*=0$ in the eq.~(\ref{powers}) for the power spectrum, 
which is equivalent to 
assume\footnote{Nevertheless, we have also studied the effect of including fully thermalized
inflaton perturbations in our results, e.g., by assuming $n_*$ takes the form of
a Bose-Einstein distribution~\cite{Ramos:2013nsa} and verified that all of our results
do not change qualitatively.}
$\Gamma/H \ll 1$. In terms of the microscopic model leading to the present WI dynamics, e.g.
through the axion-like interaction with the gauge fields, this can be achieved for sufficiently small gauge couplings and axion constant $f$. {}Finally, we also need to obtain the multiplicative function $G(Q)$ in eq.~(\ref{powers}). Existing functions, which are
obtained by fitting the numerical data for the perturbations in WI 
(see, e.g. refs.~\cite{Bastero-Gil:2011rva,Bartrum:2013fia,Das:2022ubr}) are valid in general
for $n_* \sim n_{\rm BE}$, i.e. assuming thermalization, $\Gamma > H$, and in which case $n_*$ is approximated to a Bose-Einstein distribution~\cite{Bastero-Gil:2017yzb}.
Here, we generate the appropriate data for $G(Q)$ for the model and dissipation coefficients studied by explicitly evolving the perturbation equations
in WI using the recently released \texttt{WI2easy} code for studying WI dynamics in general~\cite{Rodrigues:2025neh}. 
We refer the interested reader to ref.~\cite{Rodrigues:2025neh} for details.
All our numerical results were obtained with the help of \texttt{WI2easy}.
In fig.~\ref{fig1}, we show the numerically generated results for $G(Q)$ for the two cases
of dissipation coefficient considered here. Note that the results for $G(Q)$ are insensitive to the form of the potential, although it is strongly dependent on the form of
the dissipation coefficient (see ref.~\cite{Rodrigues:2025neh}). The numerical data for $G(Q)$ is then interpolated using a spline method and used in the spectrum eq.~(\ref{powers}) to derive the
numerical results shown in the next section. 

\section{Numerical Results}
\label{sec4}

Keeping in mind the conditions and constraints discussed in the previous section, here we present our main numerical results when considering the potential eq.~(\ref{pot}).
As in ref.~\cite{Daido:2017wwb}, we assume  in the potential eq.~(\ref{pot}) the parameters
$\kappa =1$, $n=3$ and $\theta \ll 1$. The explicit value of $\theta$ is adjusted according to the dissipation
ratio $Q$ considered such that the spectral tilt $n_s$ will fall close to the observational value $n_s \simeq 0.965$
from the Planck data\footnote{We could also perform the analysis considering the most recent data results for $n_s$
obtained from the Atacama
Cosmology Telescope (ACT)~\cite{ACT:2025tim}, like done recently in ref.~\cite{Berera:2025vsu} in the context of
WI. But this only affects our results very mildly.}.
We also observe that the dynamics is throughout in the WI regime, i.e., it satisfies $T/H > 1$, starting from the moment
the observable quantities are computed (when the relevant CMB scales cross the Hubble radius), 
which happens around 60 $e$-folds before the end of inflation.
Generically, this condition of having  $T/H > 1$ requires that $Q \gtrsim 10^{-6}$ at the moment
the relevant scales cross the Hubble radius.

As already mentioned, we consider the two cases of dissipation coefficients, with cubic and linear dependencies on the temperature,
as motivated by eqs.~(\ref{UpscubicT}) and (\ref{UpslinearT}), respectively.
{}For a nonthermalized
inflaton (see previous section) and for a thermal bath of pure gauge particles,
the effective number of degrees of freedom is taken to be given by $g_* = 2 (N_c^2-1)$ where we are considering $N_c=3$.

To determine the point ($N_*$) where the relevant pivot scale crosses the Hubble radius
during inflation, we use~\cite{Das:2020xmh}
\begin{equation}
\frac{k_*}{a_0 H_0} = e^{-N_*} \left[ \frac{43}{11 g_s(T_{\rm end})}
  \right]^{1/3} \frac{T_0}{T_{\rm end}} \frac{H_*}{H_0}  \frac{a_{\rm
    end}}{a_{\rm reh}},
\label{N*}
\end{equation}
where $g_s(T_{\rm end})$ in the entropy number of degrees of freedom at the end of inflation, taken here to be given
by $g_*$ for the models under consideration, $k_* = 0.05/{\rm Mpc}$
is the typical pivot scale and which we also consider here, $a_0=1$ is the scale factor today, $H_0=67.66\, {\rm km}\, s^{-1}
{\rm Mpc}^{-1}$ (from the Planck Collaboration~\cite{Aghanim:2018eyx},
  TT,TE,EE-lowE+lensing+BAO 68$\%$ limits) and $T_0 = 2.725\, {\rm K} \simeq 2.349
\times 10^{-13}\, {\rm GeV}$ is the present value of the CMB temperature. In the above equation, 
$a_{\rm end}/a_{\rm reh}$ gives the duration (in $e$-folds) lasting from the end of inflation until the beginning
of the radiation dominated phase (determined by solving the background equations and tracking the point where 
the equation of state after WI becomes approximately 1/3). {}Finally, the normalization of the potential $V_0$  
is determined by assuming that the scalar spectrum has the 
amplitude $\ln\left(10^{10} P_{{\cal R}} \right) \simeq 3.047$ (from the Planck Collaboration~\cite{Aghanim:2018eyx}
TT,TE,EE-lowE+lensing+BAO 68$\%$ limits), at the pivot scale $k_*$.

\subsection{Linear dissipation coefficient}

\begin{table*}[!htpb]
\centering
\caption{Parameters and results for $r$, $n_s$ for the dissipation coefficient eq.~(\ref{linearT}), in the case of $n=3$ and $\kappa=1$ in the potential eq.~(\ref{pot}).}
\setlength{\tabcolsep}{2.5pt}
\footnotesize
\begin{tabular}{ccccccccccc}
\hline \hline
$f_a/M_{\rm Pl}$ & $\theta$ & $V_0/M_{\rm Pl}^4$ &$Q_*$ & $\phi_*/M_{\rm Pl}$ & $\dot\phi_*/M_{\rm Pl}^2$ & $\rho_{r,*}/M_{\rm Pl}^4$ & $C_\Upsilon$ & $n_s$ & $r$ & $N_{*}$ \\ \hline 
$0.5$ &  $2.5\times 10^{-4}$ &   $2.18\times 10^{-14}$   & $9.96\times 10^{-7}$ & 0.0216 & $4.55\times 10^{-11}$ & $1.53 \times 10^{-27}$ & $2.60 \times 10^{-6}$ & 0.9670 & $1.24\times 10^{-6}$  & $55.6$      \\
$0.75$ &  $8.0\times 10^{-4}$ &   $1.08\times 10^{-13}$  &  $9.96\times 10^{-7}$ & 0.0493 & $2.21\times 10^{-10}$ & $3.62 \times 10^{-26}$ & $2.62 \times 10^{-6}$ & 0.9663 & $6.14\times 10^{-5}$  & $56.1$      \\
$1.0$ &  $1.9\times 10^{-3}$ &   $3.33\times 10^{-13}$ &  $9.96 \times 10^{-7}$ & 0.0878 & $6.94\times 10^{-10}$ & $3.56 \times 10^{-25}$ & $2.60 \times 10^{-6}$ & 0.9663 & $1.90\times 10^{-5}$  & $56.4$      \\   \hline \hline
\label{tab1}
\end{tabular}
\end{table*}

Let us first present the results for the case of a linear in the temperature dissipation coefficient, 
\begin{equation}
\Upsilon = C_\Upsilon T,
\label{linearT}
\end{equation}
where $C_\Upsilon$ is a dimensionless constant (for example, as given by the coefficient in eq.~(\ref{UpslinearT})).
We analyze the cases where the axion decay constant $f_a$ in the potential eq.~(\ref{pot}) can assume the
values $f_a = 0.5 M_{\rm Pl},\; 0.75 M_{\rm Pl}$ and $ 1 M_{\rm Pl}$. The relevant parameters and quantities
evaluated in these three cases, when solving the background equations of WI (\ref{eqphi}) and (\ref{eqrhor})
and from the power spectrum eq.~(\ref{powers}) are summarized in Table~\ref{tab1}.

\begin{center}
\begin{figure}[!bth]
\subfigure[Temperature over Hubble ratio]{\includegraphics[width=7.cm]{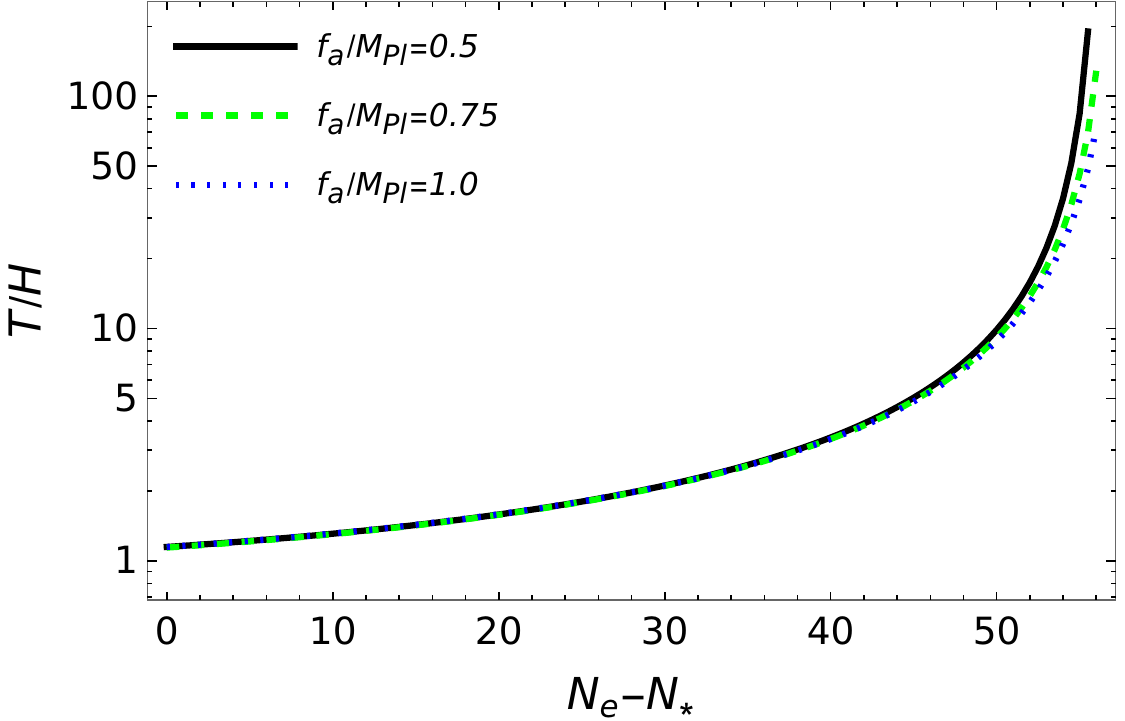}}
\subfigure[Dissipation ratio $Q$]{\includegraphics[width=7.cm]{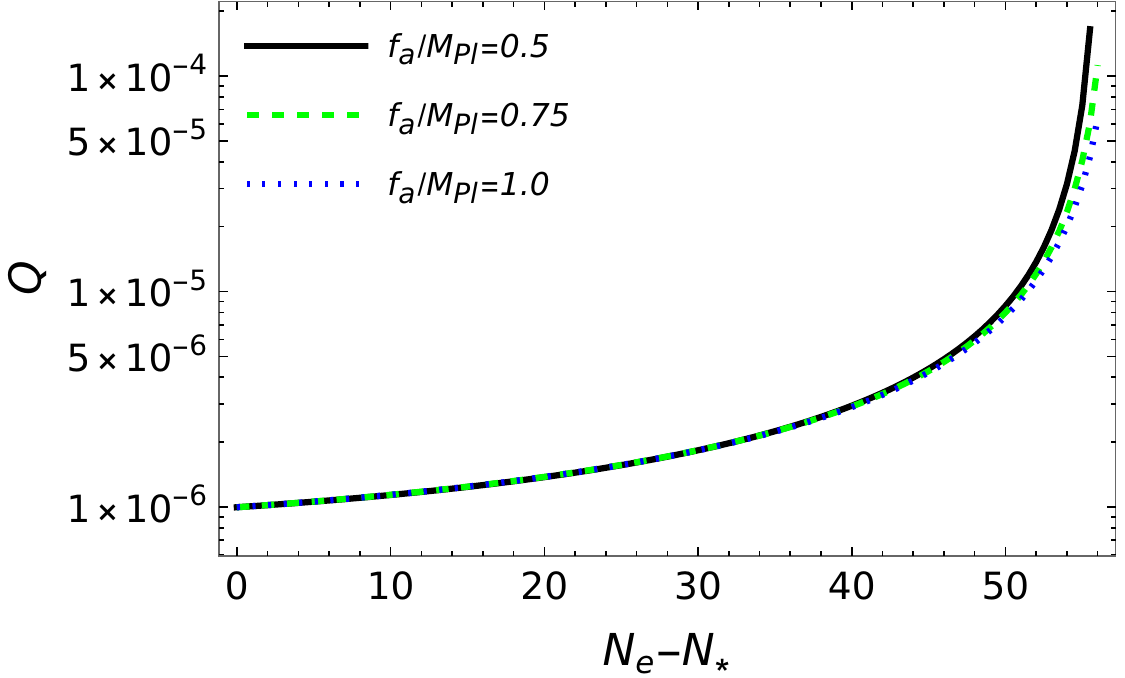}}
\subfigure[Slow-roll coefficient $\epsilon_H$]{\includegraphics[width=7.cm]{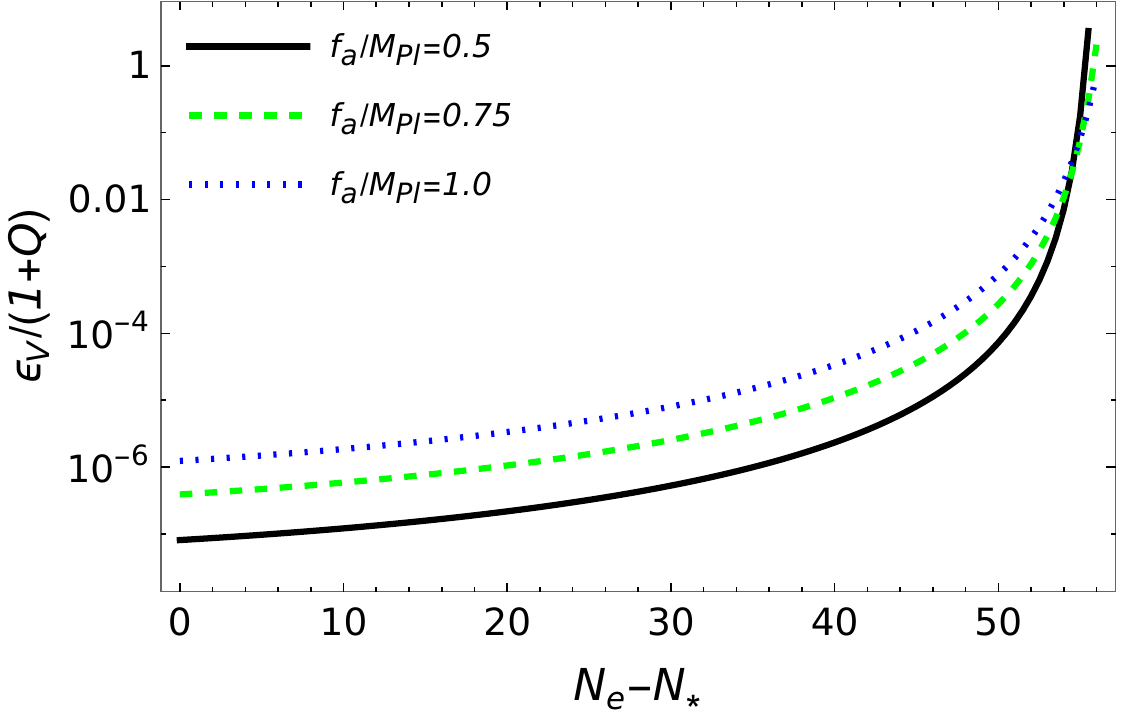}}
\subfigure[Slow-roll coefficient $\eta$]{\includegraphics[width=7.cm]{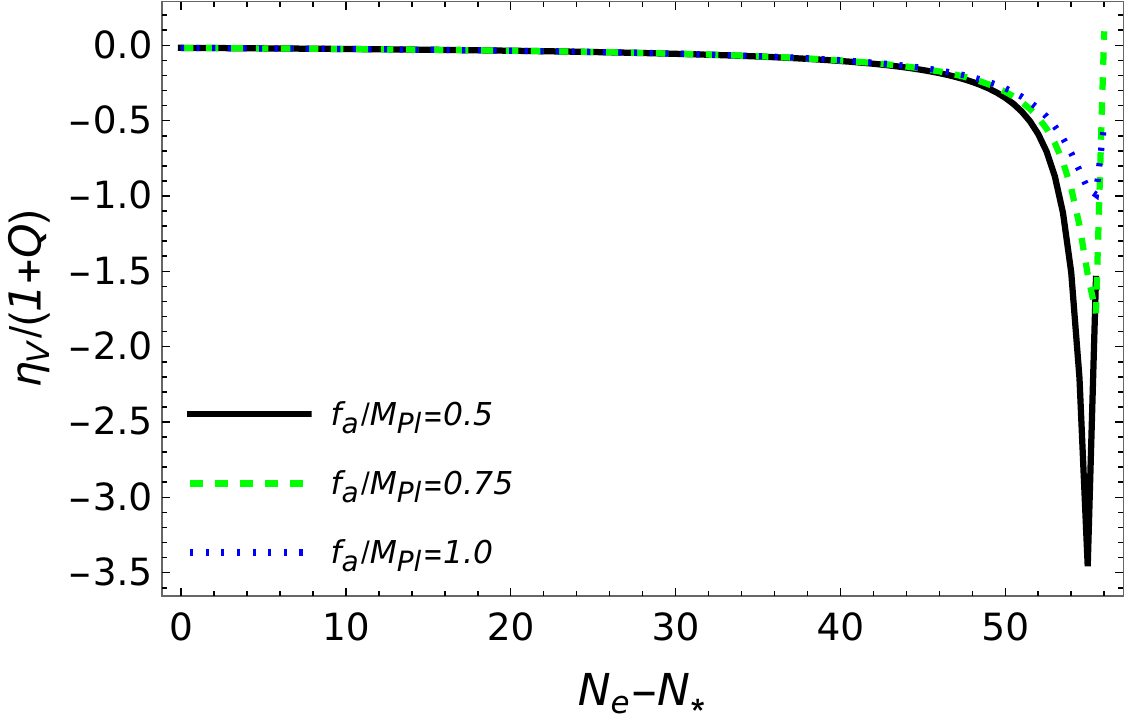}}
\caption{Main background quantities as a function of the number of $e$-folds 
for the case of the dissipation coefficient  eq.~(\ref{linearT}).}
\label{fig2}
\end{figure}
\end{center}

\begin{center}
\begin{figure*}[!bth]
\centerline{\includegraphics[width=7.5cm]{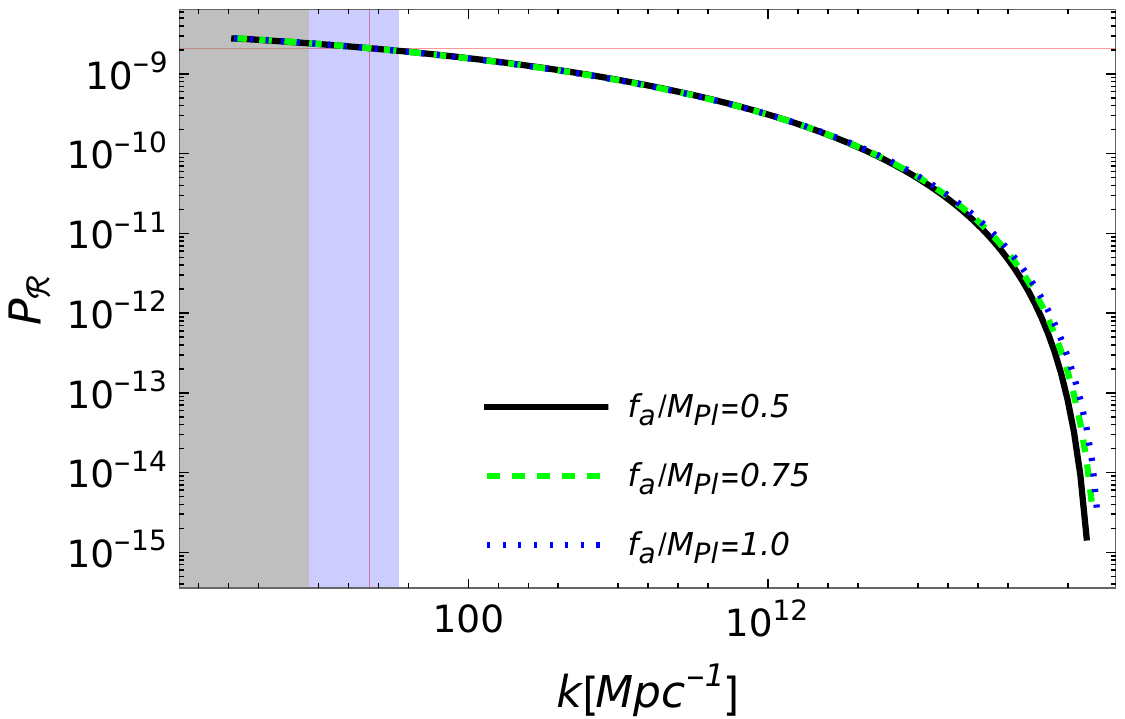}}
\caption{Power spectrum as a function of the scale for the case of the dissipation coefficient  eq.~(\ref{linearT}).
The left light blue region indicates the range of comoving modes values accessible through the current CMB data, 
$k_{\rm CMB} \in [0.0005,0.5]{\rm Mpc}^{-1}$.
The most to the left gray region indicates the range of comoving modes outside the horizon.}
\label{fig3}
\end{figure*}
\end{center}

In {}fig.~\ref{fig2} we show the evolution for the main background quantities, namely $T/H$, which confirms that we are in WI regime ($T/H > 1$)
throughout the dynamics, the evolution of the dissipation ratio $Q$ and also the evolution of the slow-roll coefficients in WI.
In {}Fig~\ref{fig3}
we show the power spectrum eq.~(\ref{powers}),  as a function of the scale, for the cases of the dissipation coefficient eq.~(\ref{linearT}) for the three cases
of values $f_a$ considered.

\subsection{Cubic dissipation coefficient}

\begin{table*}[!htpb]
\centering
\caption{Parameters and results for $r$, $n_s$ for the cubic in $T$ dissipation coefficient $\Upsilon=C_\Upsilon T^3/M_{\rm Pl}^2$, in the case of $n=3$ and $\kappa=1$ in the potential eq.~(\ref{pot}).}
\setlength{\tabcolsep}{2.5pt}
\footnotesize
\begin{tabular}{ccccccccccc}
\hline \hline
$f_a/M_{\rm Pl}$ & $\theta$ & $V_0/M_{\rm Pl}^4$ &$Q_*$ & $\phi_*/M_{\rm Pl}$ & $\dot\phi_*/M_{\rm Pl}^2$ & $\rho_{r,*}/M_{\rm Pl}^4$ & $C_\Upsilon$ & $n_s$ & $r$ & $N_{*}$ \\ \hline 
$0.5$ &  $1.5\times 10^{-3}$ &   $4.26\times 10^{-13}$   & $9.11\times 10^{-7}$ & 0.0229 & $9.14\times 10^{-10}$ & $5.54 \times 10^{-25}$ & $7.43 \times 10^6$ & 0.9660 & $2.43\times 10^{-5}$  & $55.5$      \\
$0.75$ &  $3.8\times 10^{-3}$ &   $1.33\times 10^{-12}$  &  $9.04\times 10^{-7}$ & 0.0527 & $2.81\times 10^{-9}$ & $5.18 \times 10^{-24}$ & $2.43 \times 10^6$ & 0.9640 & $7.56\times 10^{-5}$  & $55.7$      \\
$1.0$ &  $7.2\times 10^{-3}$ &   $2.91\times 10^{-12}$ &  $8.97 \times 10^{-7}$ & 0.0965 & $6.14\times 10^{-9}$ & $2.45 \times 10^{-23}$ & $1.12 \times 10^6$ & 0.9617 & $1.66\times 10^{-4}$  & $56.0$      \\   \hline \hline
\label{tab2}
\end{tabular}
\end{table*}

We now consider the results for the case of the dissipation coefficient of the form of eq.~(\ref{UpscubicT}), 
i.e.,
\begin{equation}
\Upsilon = C_\Upsilon \frac{T^3}{M_{\rm Pl}^2},
\label{cubicT}
\end{equation}
where $C_\Upsilon$ is also a dimensionless constant (for example, as given by the coefficient in eq.~(\ref{UpscubicT})).
We once again analyze the same cases studied in the previous case of dissipation coefficient, by considering
the axion decay constant $f_a$ in the potential eq.~(\ref{pot}) to have the
values $f_a = 0.5 M_{\rm Pl},\; 0.75 M_{\rm Pl}$ and $ 1 M_{\rm Pl}$. The relevant parameters and quantities 
derived for this
dissipation coefficient are now summarized in the Table~\ref{tab2}.

\begin{center}
\begin{figure}[!bth]
\subfigure[Temperature over Hubble ratio]{\includegraphics[width=7.cm]{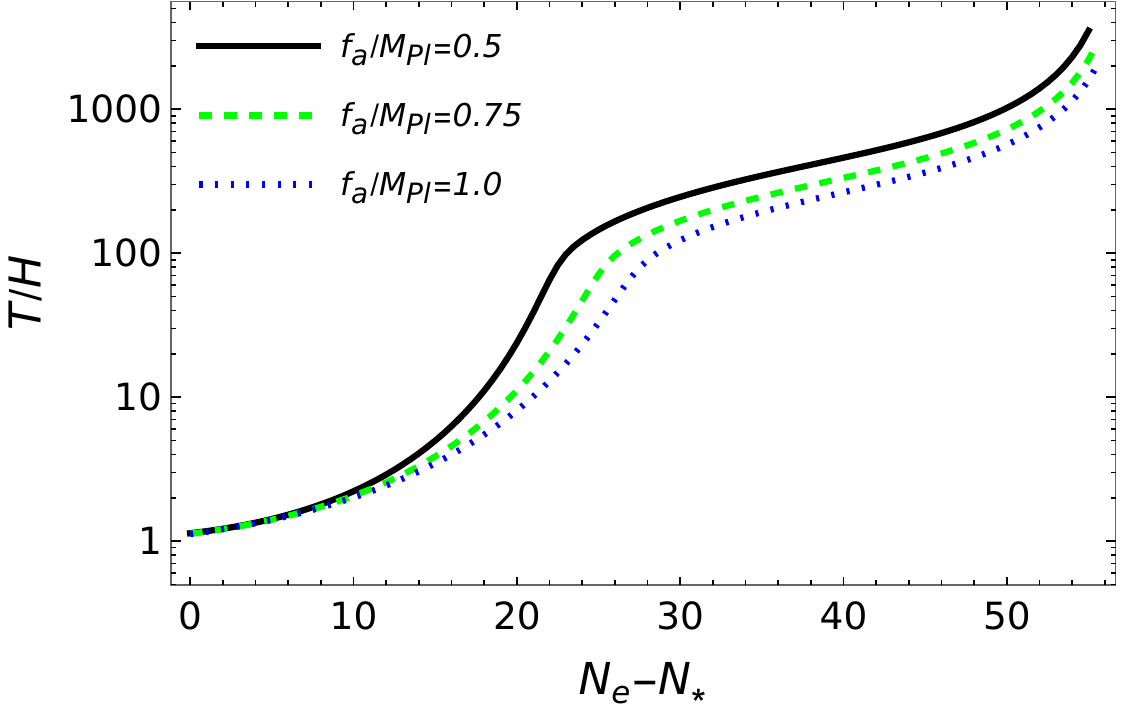}}
\subfigure[Dissipation ratio $Q$]{\includegraphics[width=7.cm]{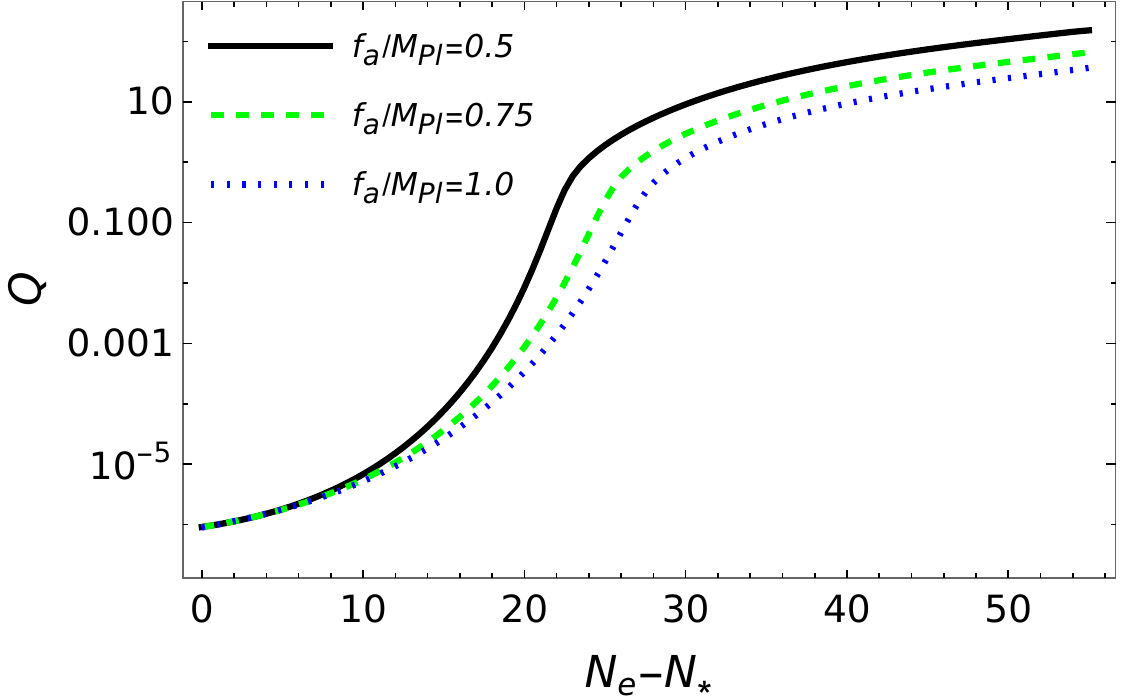}}
\subfigure[Slow-roll coefficient $\epsilon_H$]{\includegraphics[width=7.cm]{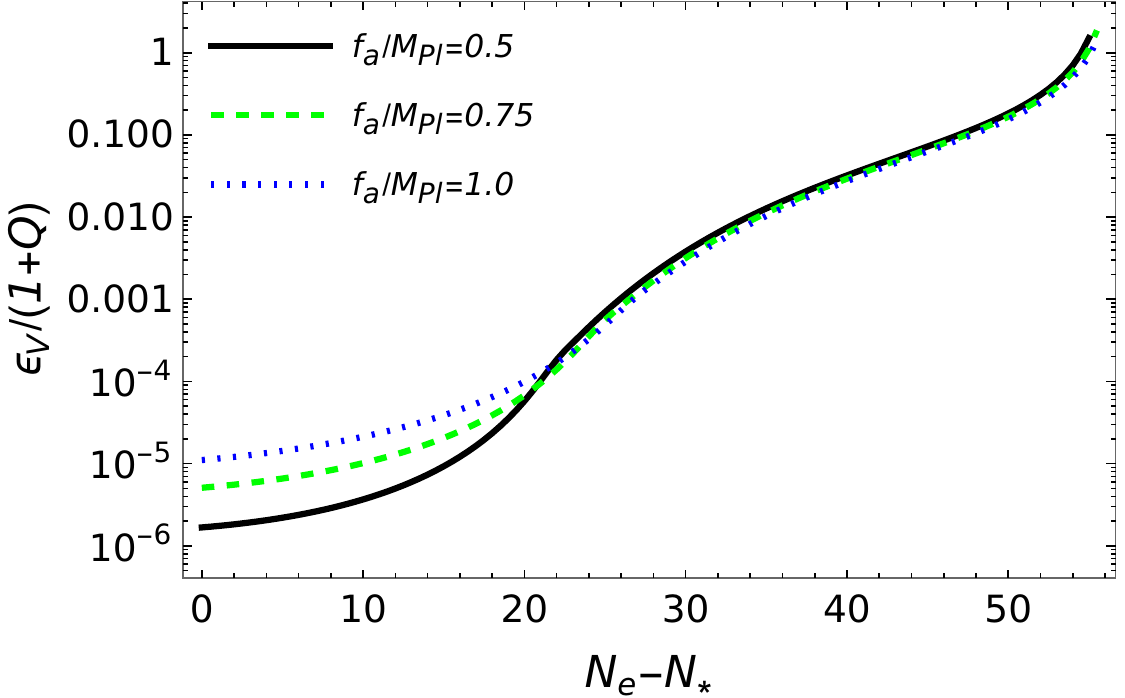}}
\subfigure[Slow-roll coefficient $\eta$]{\includegraphics[width=7.cm]{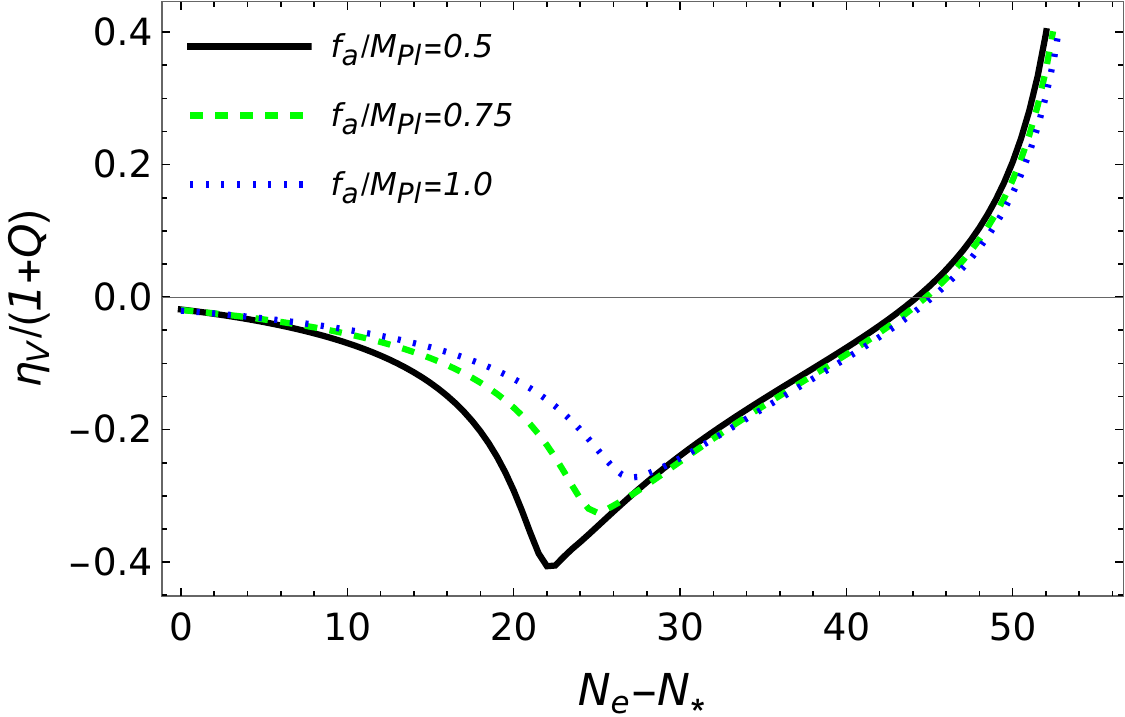}}
\caption{Main background quantities as a function of the number of $e$-folds 
for the case of the dissipation coefficient  eq.~(\ref{cubicT}).}
\label{fig4}
\end{figure}
\end{center}

\begin{center}
\begin{figure*}[!bth]
\centerline{\includegraphics[width=7.5cm]{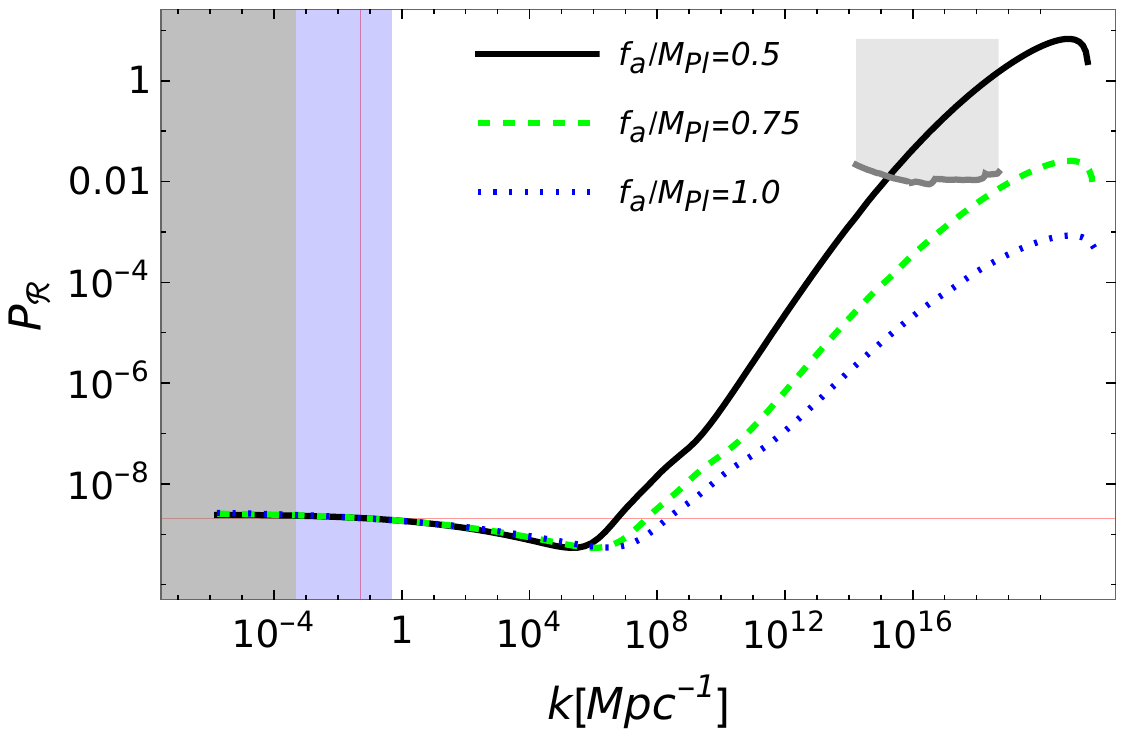}}
\caption{Power spectrum as a function of the scale for the case of the dissipation coefficient  eq.~(\ref{cubicT}).
The top gray region is excluded due to nucleosynthesis bounds~\cite{Carr:2020gox}.
The left light blue region indicates the range of comoving modes values accessible through the current CMB data, 
$k_{\rm CMB} \in [0.0005,0.5]{\rm Mpc}^{-1}$.
The most to the left gray region indicates the range of comoving modes outside the horizon.}
\label{fig5}
\end{figure*}
\end{center}

In {}fig.~\ref{fig4} we show the evolution for the main background quantities, namely $T/H$, which also confirms that we are in WI regime ($T/H > 1$)
throughout the dynamics, the evolution of the dissipation ratio $Q$ and for the slow-roll coefficients in WI. In {}fig.~\ref{fig5}
we show the power spectrum eq.~(\ref{powers}) as a function of the scale that is obtained for the case of a cubic in the temperature
dissipation coefficient in the multi-natural WI model.

\subsection{Discussion of the results}

{}From {}figs.~\ref{fig3} and \ref{fig5}, we see that while in the case of the dissipation coefficient
which has a linear dependence on the temperature, the power spectrum always evolves such that to
have less power at small scales, the situation inverts in the case of the dissipation
coefficient with a cubic dependence on the temperature.
In the cubic dissipation case, the power spectrum tends to grow significantly at the end
of inflation. This is a consequence of both the form of the dissipation coefficient and the
form of the potential, where inflation occurs mostly in the concave section of $V(\phi)$.
We could anticipate this expectation based on the analytical expressions presented in the
previous section (see the discussion in the paragraph below eq.~(\ref{b4})) and how they enter
in the expression of the WI power spectrum eq.~(\ref{powers}).
As discussed, in the case of a dissipation coefficient with a cubic in the temperature dependence
and for inflaton potentials that can have a concave part where inflation happens, like in the model
under study, the power spectrum can exhibit rapid growth.
In particular, linear cosmological perturbations can break down when the power spectrum
goes above one. {}For the parameters assumed, for an axion decay constant $ f_a \lesssim 0.75M_{\rm Pl}$, 
this can happen already
for a very small dissipation ratio, $Q \gtrsim 10^{-6}$, and will become amplified for higher
values of $Q$. Changing the potential parameters, in particular making $\theta$ smaller, 
to obtain a correct value for $n_s$ we would need to increase
$Q$ initially, which would already put it at a value above the threshold leading
to  $P_{\cal R} > 1$ towards the end of inflation.
{}From {}fig.~\ref{fig4}, we see that the behavior of $\eta_{V}$, $Q$ and $T/H$ change drastically around 
the number of {\it e}-folds for which $Q \sim 1$.
In fact, though $\eta_{V}$ decreases before $Q \sim 1$, it suddenly starts to increase after the 
turning point. Indeed, we can trace the point where the power spectrum shown in {}fig.\,\ref{fig5} stops losing power and transitions
to a growing behavior exactly at this point.
This feature has been observed in other WI models with
different inflaton 
potentials~\cite{Arya:2019wck,Bastero-Gil:2021fac,Correa:2022ngq}.

{}Finally, we discuss the observational compatibility of the results obtained.
There exist observational bounds on the fraction of primordial black holes (PBH) from 
their evaporation~\cite{Carr:2009jm,Carr:2020gox}.
The results in {}fig.\,\ref{fig5} show that the power spectrum peaks at very small scales,
close to the Big Bang Nucleosynthesis (BBN) scales.
However, in general, it is not straightforward how to recast the bounds into those on 
the curvature perturbations.
In fact, there are several uncertainties, such 
as the density-contrast threshold~\cite{Carr:2020gox,Carr:2021bzv},
which can change the final result exponentially, 
when one calculates the fraction of PBH 
from a given power spectrum of curvature perturbations, and vice versa.
Therefore, currently, we would not have a concrete way to evaluate the BBN bound on the scalar power spectrum through the primordial black hole evaporation.
In {}fig.\,\ref{fig5}, 
we just compare an example of a BBN bound (top gray region in the plot) in the small scale power spectrum obtained 
in ref.~\cite{Carr:2020gox} with
our scalar power spectra, just as a benchmark.
We can see that the case of $f_a=0.5M_{\rm Pl}$ seems to be incompatible with the BBN bound, while 
$f_a=0.75M_{\rm Pl}$ (marginally) and $f_a=1.0M_{\rm Pl}$ may be compatible with it.

{}For the linear dissipation case, the power spectrum remains well under control throughout
the evolution up to the end of inflation. Typically, to obtain a consistent value for $n_s$
also implies values for the dissipation ratio $Q_*$ that are in the weak regime of WI.
In this weak regime, $Q\ll 1$, there is no rapid growth of the power spectrum, in contrast to the cubic dissipation case.

In both cases of dissipation coefficients, in principle we could lower the axion decay constant $f_a$ in the potential,
but at the cost of further decreasing $Q_*$ initially. Here in the multi-natural axion-type of
potential eq.~(\ref{pot}), we see a difficulty in achieving sub-Planckian axion decay 
constants and a consistent regime of WI, analogous to what also happens in vanilla single-axion cases~\cite{Montefalcone:2022jfw,Zell:2024vfn}.

\section{Induced gravitational waves}
\label{GWsection}

In the previous section, we saw that scalar fluctuations can grow on small scales for the case of the cubic dissipation 
coefficient, i.e. eq.\,(\ref{cubicT}).
We also showed that there exist several parameter sets which give rise to the scalar power spectrum which may be compatible with the BBN bound.
Interestingly, the magnitude of the power spectrum can still be large, of the order $\mathcal{O}(0.01)$ around the end of inflation.
The tensor parts of perturbations do not interact with the scalar parts of perturbations at a linear level in an isotropic background universe. 
Thus, tensor and scalar fluctuations interact with each other only at the non-linear level~\cite{Baumann:2007zm,Ananda:2006af}.
This is the reason why one can consider their evolution independently, since non-linear interactions are negligible in practice.
However, when we have large scalar fluctuations, the non-linear interaction between tensor and scalar perturbations would not be negligible 
anymore.
In fact, large scalar fluctuations can produce abundant gravitational waves~\cite{Saito:2008jc}.
In this section, we calculate the scalar-induced gravitational waves
on small scales in the case of the cubic dissipation coefficient.

\begin{center}
\begin{figure}[!bth]
\centerline{\includegraphics[width=8.cm]{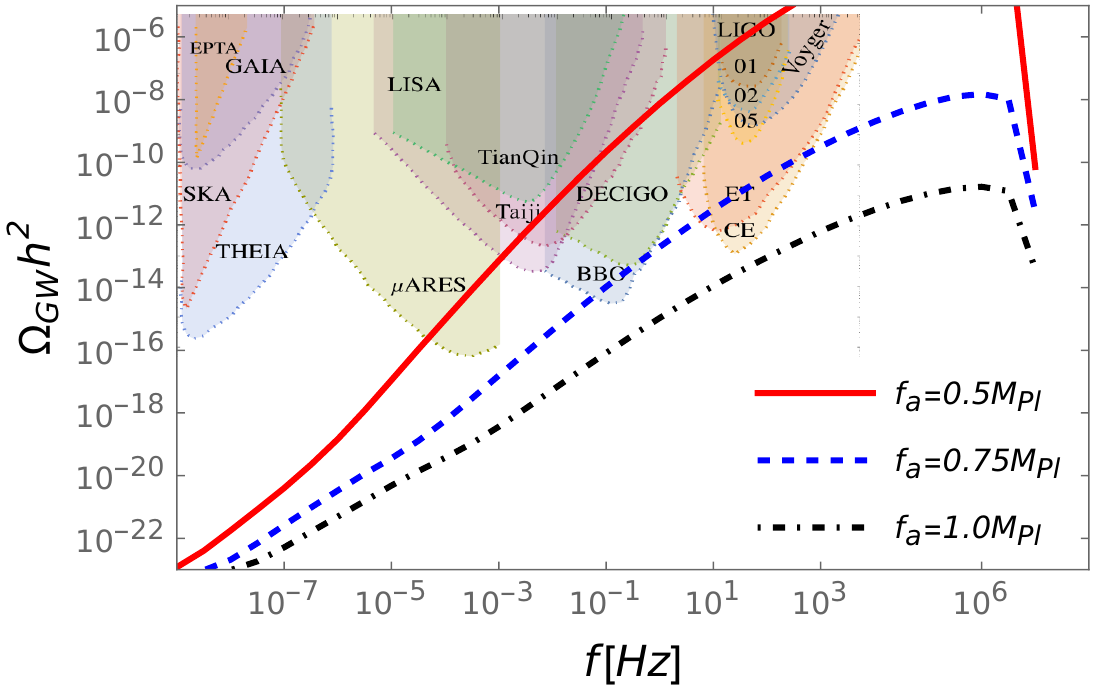}}
\caption{The spectrum of produced secondary gravitational waves as a function of frequency for the case
of the dissipation coefficient eq.~\ref{cubicT}. The top shaded regions show the
power-law integrated sensitivity for the current and future gravitational wave detectors
(regions taken from {}fig.1 of ref.~\cite{Roshan:2024qnv}).}
\label{fig6}
\end{figure}
\end{center}

WI ends when the radiation energy density surpasses that of the inflaton, namely the radiation dominated era begins 
after the inflation ends\footnote{In practice, it still can take around one to two {\it e}-folds
of expansion after the end of WI before radiation energy density becomes dominant~\cite{Das:2020lut}.}.
Then, super-horizon modes of fluctuations created around the end of inflation experience re-entering in the radiation-dominated era. 
It is known that an approximated analytical expression of the power spectrum of gravitational waves produced by nonlinear scalar perturbations
in the radiation dominated era is given by~\cite{Kohri:2018awv}
\begin{equation}
  P_{h}(k) \simeq 4 
                 \int_{0}^{\infty} dp \int^{k+p}_{|k-p|} dq 
                 \frac{p^{2}}{k^{2} q^{2}} 
                 \left[ 1-\left( \frac{k^{2} + p^{2} -q^{2} }{2 k p} \right)^{2}  \right]^{2}
                 I^{2}(p,q,\eta) P_{R}(p) P_{R}(q),
\label{Phspec}
\end{equation}
where $\eta$ is the conformal time and  
\begin{eqnarray}
 I^{2}(p,q,\eta) &\simeq & \frac{1}{2}\left[ \frac{3(p^{2}+q^{2}-3k^{2})}{4p^{2}q^{2}k\eta} \right]^{2}
                 \left\{ \left[ 4- \left(\frac{p}{q} + \frac{q}{p} - \frac{3k^{2}}{pq}\right) 
                        \ln\left|\frac{3k^{2}-(p+q)^{2}}{3k^{2}-(p-q)^{2}} \right|  \right]^{2}              
                         \right. \nonumber\\
&+&\left. \pi^{2}\left(\frac{p}{q} + \frac{q}{p} - \frac{3k^{2}}{pq}\right)^{2}
                          \theta(p+q-\sqrt{3}k)  \right\} .
\end{eqnarray}
The power spectrum eq.~(\ref{Phspec}) is related to the energy density parameter as
\begin{equation}
  \Omega_{{\rm GW}}(k,\eta) = \frac{k^{2}}{24a(\eta)^2H(\eta)^{2}}  P_{h}(k) , 
\end{equation}
where $H(\eta)$ represents the Hubble parameter.
The present spectra of $\Omega_{{\rm GW}}$ induced by
the scalar power spectra given in fig.~\ref{fig5} are shown in {}fig.~\ref{fig6}, 
while assuming that
the radiation dominant phase continues until the radiation-matter equality.
One can see that the produced secondary gravitational waves
can be tested with future gravitational wave detectors 
such as DECIGO, BBO, ET, and CE.

It should be mentioned that an additional peak in a gravitational wave spectrum could appear at a lower 
frequency if a certain fraction of light PBH were produced~\cite{Inomata:2020lmk,Domenech:2020ssp,Domenech:2024wao} and a 
dominant PBH phase appeared.
Although this possibility is interesting, we do not consider such a possibility.
This is because there are several uncertainties, which can change the
result exponentially, in the calculation of the fraction of primordial black holes from 
a given scalar power spectrum as discussed in the previous section.

\section{Conclusion}  
\label{conclusions}

In this paper, we studied multi-natural inflation in the context of WI.
We studied the warm multi-natural inflation with dissipation coefficients that have either a linear or a cubic dependence on the
temperature. These types of dissipation coefficients are well motivated in WI models where the inflaton is a pseudo-Goldstone
scalar field, and they appear as limiting cases of dissipation from sphaleron decay from gauge and fermion interactions.
It turned out that both cases could result in the spectral tilt and the tensor-to-scalar ratio, which are
compatible with current observations.
We also chose to closely follow the motivation for multi-natural inflation in producing both consistent inflationary observables and, 
eventually, a small mass for the inflaton at the end of inflation, such that the inflaton could also serve as dark matter. 
Therefore, we have considered only the case of an odd value for $n$ in the potential eq.~(\ref{pot}). However, WI would also work
similarly for even values of $n$ leading to qualitatively similar results than the ones we have obtained.
It is worth noting that in WI, as opposite to the case of CI, we found that a smaller axion decay constant leads to a larger enhancement of 
the power spectrum. This presents a challenge for the inflaton to also serve as dark matter in the scenarios proposed e.g. in ref.~\cite{Daido:2017wwb},
as far as its implementation in WI is concerned.

As for the two cases of the dissipation coefficient that we have studied, we found that in 
the case of the cubic dissipation coefficient, in general, the curvature perturbation starts to increase
suddenly when a transition from a weak
to a strong dissipation regime occurs at the middle stages of the inflation dynamics.
We have analyzed this behavior of the power spectrum for the model studied and its implications
for both PBH formation and gravitational waves.
We have calculated the scalar-induced gravitational waves from the large curvature perturbations on
small scales.
Interestingly, it was shown that gravitational waves may be detectable with future gravitational wave
detectors such as DECIGO and ET.
This indicates that the warm multi-natural inflation scenario is testable with gravitational wave 
observations in principle.
Moreover, since the produced gravitational wave spectra have peaks around MHz, the development of new detection methods
for high-frequency gravitational waves is
encouraged~\cite{Aggarwal:2025noe,Matsuo:2025blj,Ito:2023nkq,Ito:2023bnu,Ito:2023fcr,Ito:2022rxn,Ito:2020wxi,Ito:2019wcb}.

It is worth to notice that while enhancements of the power spectrum in CI are typically achieved by tweaking the inflaton potential 
and adding features to it in some unnatural ways, in WI this is not necessary. {}For the type of potential we have considered, the enhancement of
the power spectrum occurs naturally, with the dynamics already allowing the
spectrum to change from red-tilted to blue-tilted. As for possible signatures that could distinguish this model from 
others, we would say that they would be the same signatures that would
distinguish CI from WI. The most prominent one would be the differences in non-Gaussianities in the two scenarios (see ref.~\cite{Bastero-Gil:2014raa}).
WI also does not satisfy the consistency relation between the tensor-to-scalar ratio and the tilt of the tensor spectrum, $r=-8n_t$, as observed in CI. Typically, we find that in WI,  $r< 8|n_t|$ (see, e.g. refs.~\cite{Bartrum:2013fia,Benetti:2016jhf}). Another
distinctive feature of WI is also the possibility of the generation of matter isocurvature perturbations, due to the dissipative effects during inflation,
that are fully anti-correlated with the dominant adiabatic curvature perturbations~\cite{Bastero-Gil:2014oga}. 

It should be mentioned that although we have not taken into account the non-Gaussianity 
of the curvature perturbations in the calculation of the induced gravitational waves,
such an effect could change the tensor power spectrum slightly~\cite{Nakama:2016gzw,Cai:2018dig,Unal:2018yaa}.
The non-Gaussianity of the curvature perturbations also can affect the fraction of created PBH, though its contribution is expected to be small~\cite{Saito:2008em}.
Even apart from the effect of the non-Gaussianity, there still exist several uncertainties which can alter the
result exponentially, like in the calculation of the fraction of PBH from 
a given scalar power spectrum, such 
as the density-contrast threshold~\cite{Carr:2020gox,Carr:2021bzv}.
Therefore, for benchmark purposes, we just gave a comparison between an example of the BBN bound on the small scale power spectrum given 
in~\cite{Carr:2020gox} and
our scalar power spectra which have peaks right before the end of the inflation in the case of the cubic dissipation coefficient.
Related to this point, we also assumed that there was no dominant PBH epoch after
inflation, though
it appears inevitably if certain amount of PBH are produced~\cite{Domenech:2020ssp,Domenech:2024wao}.
Interestingly, if this is the case, not only the current  gravitational wave spectrum is deformed, but also 
an additional peak could appear in the spectrum~\cite{Inomata:2020lmk,Domenech:2020ssp,Domenech:2024wao}.
We hope these aspects will be revisited elsewhere in the future.

\appendix

\section{Spectral tilt in the multi-natural WI model}
\label{appA}

In this appendix we derive approximate expressions for the spectral tilt
$n_s$ which helps to understand the behavior shown in figs.~\ref{fig3} and \ref{fig5}.

{}From eq.~(\ref{powers}), the scalar spectral index $n_s$ can be
calculated at the horizon crossing ($k_*$) as
\begin{equation}
n_s -1 =   \frac{d \ln P_{{\cal R}} }{d \ln k}\Bigr|_{k\to k_*} ,
\label{eq:n}
\end{equation}
where, at Hubble radius crossing scale $k_*=aH$, we can write $d\ln
k=(d\ln k/dN_e) dN_e$,  $N_e=\ln a$ being the number of $e$-folds, then
\begin{equation}
\frac{d\ln k}{dN_e}\approx  1-\epsilon_V/(1+Q).
\label{dlnk}
\end{equation}

{}From eq.~(\ref{powers}), where we here consider $n_*=0$ (see discussion following eq.~(\ref{rate})), 
together with (\ref{dlnk}), we then obtain for
$n_s$ the result
\begin{eqnarray}
n_s-1&=& \frac{-6 \epsilon_V+2
    \eta_V}{1+Q-\epsilon_V} + \frac{d \ln Q}{dN} \frac{1+Q}{1+Q-\epsilon_V}{\cal A}(Q)
\nonumber \\
&+&  \frac{d \ln Q}{dN}     
\frac{-3+(3-2 \pi ) Q+2 \pi  Q^2 \left(3+2 \sqrt{9+12 \pi  Q}\frac{T}{H}\right)}{\left[3+2 \pi  Q \left(2+\sqrt{9+12 \pi  Q}\frac{ T}{H}
\right)\right]
\left(1+Q-\epsilon_V\right)}
\nonumber \\
&+& \frac{d 
    \left(\frac{T}{H}\right)}{dN} \frac{2 \pi  Q (1+Q) \sqrt{9+12 \pi  Q} }{\left[3+2 \pi  Q 
\left(2+\sqrt{9+12 \pi  Q}\frac{ T}{H}\right)\right] 
\left(1+Q-\epsilon_V\right)},
\nonumber\\
\label{nsQ}
\end{eqnarray}
with $d \ln Q/dN$ and $d \left(\frac{T}{H}\right)/dN$ given by eqs.~(\ref{b3}) and (\ref{b4}), respectively,
and the function ${\cal A}(Q)$ is defined by
\begin{eqnarray}
\mathcal{A}(Q)=\frac{3+2\pi Q}{3+4\pi Q}+Q\frac{d\ln G(Q)}{dQ}.
\label{AQ}
\end{eqnarray}
The expression (\ref{nsQ}) is a complicated expression in terms of the various parameters
of the potential and also depends on the form of the dissipation coefficient. However, we can
make some approximate expressions for the two cases of dissipation coefficients
that we have considered, given by eqs.~(\ref{linearT}) and (\ref{cubicT}). 
Using eq.~(\ref{nsQ}) together with eqs.~(\ref{b3}) and (\ref{b4}),  we find that $n_s$ in the weak dissipative
regime $Q \ll 1$ is given by
\begin{equation}
n_s\Bigr|_{Q\ll 1} -1 \simeq -6\epsilon_V+2 \eta_V
+ \left\{
\begin{array}{cc}
10 \left( \epsilon_V -\frac{\eta_V}{3} \right)Q,   & \; {\rm for\;} \Upsilon \propto T, \\
\left(26 \epsilon_V - 14\eta_V \right)Q,   & \; {\rm for\;} \Upsilon \propto T^3.
\end{array}
\right.
\label{nsQsmall}
\end{equation}
In eq.~(\ref{nsQsmall}) the first two terms on the right-hand side reproduce
the standard CI result. The last term that depends on
$Q$ is the contribution of WI. One clearly sees that in both cases of dissipation in WI, the spectrum will be driven to become blue tilted when $\epsilon_V -\eta_V/3 >0 $ in the case of the dissipation coefficient (\ref{linearT}), while that will tend
to happen for $26 \epsilon_V - 14\eta_V > 0$
in the case of the dissipation coefficient  (\ref{cubicT}). We can also estimate the value of $Q$
that the spectrum tends to transit from red to
blue tilted. As a reference point, we can focus on the value for the inflaton where both background and spectrum tend to change most,
which is expected to happen around the inflection point of the potential\footnote{We note that for the
dissipation coefficient $\Upsilon \propto T$ and for the parameters shown in table~\ref{tab1}, 
inflation tends to end close to the inflection point of the potential. {}For $\Upsilon \propto T^3$, inflation tends to proceed beyond the inflection point and ends between this point
and the minimum of the potential.}.
{}From eq.~(\ref{pot}) and for the parameters we are considering, 
the inflection point of the potential is
\begin{equation}
\phi_{\rm inf} \simeq \frac{f_a}{2} \left( \pi - \frac{\theta}{2} \right).
\label{phiinf}
\end{equation}
Hence, from eq.~(\ref{nsQsmall}), we obtain that
\begin{equation}
n_s\Bigr|_{Q\ll 1, \phi= \phi_{\rm inf}} -1 \sim 
\left\{
\begin{array}{cc}
\frac{9 M_{\rm Pl}^2}{8 f_a^2} (2 + 3 \theta) 
\left(-3+5 Q \right),   & \; {\rm for\;} \Upsilon \propto T, \\
\frac{9 M_{\rm Pl}^2}{8 f_a^2} (2 + 3 \theta) 
\left(-3+13 Q \right),   & \;{\rm for\;} \Upsilon \propto T^3.
\end{array}
\right.
\label{nsQsmallinf}
\end{equation}
{}From eq.~(\ref{nsQsmallinf}), we find that the spectrum will tend
to change from red ($n_s <1$) to blue ($n_s>1$) for $Q \gtrsim 3/5$ in the case of a dissipation
coefficient with linear temperature dependence, while for $\Upsilon \propto T^3$
the change occurs for $Q \gtrsim 3/13$.
This explains the results for the scalar of curvature power spectrum obtained in the text for each
case. {}From fig.~\ref{fig2}(b) we see that for the parameters considered in the case of a dissipation
coefficient with linear temperature dependence, $Q$ is always much smaller than the red to blue
transition value. Hence, the power spectrum for this case shown in fig.~\ref{fig3} decreases throughout
the evolution. However, for the dissipation coefficient with a cubic temperature dependence, the turn-around
point, as seen from fig.~\ref{fig4}(b) already happens around 20 and 30 e-folds in the evolution, which also
corresponds to the scales for which we see the behavior for the power spectrum changing
in fig.~\ref{fig5}. These results of the spectrum showing the change in behavior from red to blue also persist in the
strong dissipative regime $Q > 1$. Although these results were obtained by choosing $n=3$
in the potential, we have verified (although not shown here) that they also hold for other values of $n$.
{}Furthermore, from eq.~(\ref{nsQsmallinf}), we also explicitly see that
the tilt of the spectrum (either in the red or blue directions) will be enhanced the more sub-Planckian
($f_a < M_{\rm Pl}$) the axion decay constant becomes. This is fully consistent with the observed effect of changing
the value of $f_a$ also seen in figs.~\ref{fig3} and \ref{fig5}.

\acknowledgments
A.I. would like to thank Jan Tr\"{a}nkle for helpful discussions.
A.I. was in part supported by JSPS KAKENHI Grant Number JP22K14034.
R.O.R. acknowledges financial support by research grants from Conselho
Nacional de Desenvolvimento Cient\'{\i}fico e Tecnol\'ogico (CNPq),
Grant No. 307286/2021-5, and from Funda\c{c}\~ao Carlos Chagas Filho
de Amparo \`a Pesquisa do Estado do Rio de Janeiro (FAPERJ), Grant
No. E-26/201.150/2021.


\end{document}